\documentclass[prl,aps,showpacs,twocolumn,superscriptaddress,10pt]{revtex4-1}
\usepackage{amsmath}
\usepackage{amssymb}
\usepackage{epsfig}
\usepackage{bbold}

\begin{document}

\title{Morse potential derived from first principles}

\author{Raimundo N. Costa Filho} 
\email{rai@fisica.ufc.br} \affiliation{Departamento de F\'isica,
  Universidade Federal do Cear\'a, Caixa Postal 6030, Campus do Pici,
  60455-760 Fortaleza, Cear\'a, Brazil}

\author{Geov\'a Alencar}
\email{geova@fisica.ufc.br}\affiliation{Universidade Estadual do
  Cear\'a, Faculdade de Educa\c c\~ao, Ci\^encias e Letras do Sert\~ao
  Central, R. Epit\'acio Pessoa, 2554, 63.900-000 Quixad\'{a},
  Cear\'{a}, Brazil. }

\author{Bo-Sture Skagerstam }
\email{bo-sture.skagerstam@ntnu.no}\affiliation{Department of Physics,
  The Norwegian University of Science and Technology, N-7491
  Trondheim, Norway\\
  Centre for Advanced Study (CAS), Drammensveien 78, N-0271 Oslo,
  Norway}

\author{Jos\'e S. Andrade Jr.}  
\email{soares@fisica.ufc.br} \affiliation{Departamento de F\'isica,
  Universidade Federal do Cear\'a, Caixa Postal 6030, Campus do Pici,
  60455-760 Fortaleza, Cear\'a, Brazil}

\begin{abstract}
  We show that a direct connection can be drawn, based on fundamental
  quantum principles, between the Morse potential, extensively used as
  an empirical description for the atomic interaction in diatomic
  molecules, and the harmonic potential. This is conceptually achieved
  here through a non-additive translation operator, whose action leads
  to a perfect equivalence between the quantum harmonic oscillator in
  deformed space and the quantum Morse oscillator in regular space.
  In this way, our theoretical approach provides a distinctive first
  principle rationale for anharmonicity, therefore revealing a
  possible quantum origin for several related properties as, for
  example, the dissociation energy of diatomic molecules and the
  deformation of cubic metals.
\end{abstract}

\maketitle

The quantum harmonic oscillator (QHO) is certainly one of the most
celebrated paradigms in quantum mechanics. Among its several
important attributes, the QHO can be solved exactly and has been
consistently used to approximate any potential function when Taylor
expanded around their minima till second order. Moreover, the fact
that bosons can be conceptually modeled in terms of a QHO readily
explains its broad application, ranging from fundamental physics, in
the description of a quantized electromagnetic field, to condensed
matter, for vibrational properties of molecules as well as phonons in
solids. However, anharmonic potentials are very often required to
mathematically represent physical phenomena. For instance, an adequate
description of the vibrational modes in diatomic molecules must
necessarily allow for dissociation (i.e., bond breaking) of its two
bounded atomic nuclei. By increasing the energy of the molecule,
through heating for example, the system starts to vibrate till
eventually break down into two non-bounded atoms. This is an essential
feature of diatomic molecules that is not compatible with the QHO
model.
 
As originally proposed by Phillip M. Morse in 1929 \cite{Morse1929},
the so-called Morse potential provides a much better description for
the potential energy of a diatomic molecule than the QHO, being
usually written as,
\begin{equation}
\label{Morse}
V_{\rm M}(x) = D(1-e^{-\alpha x})^{2},
\end{equation}
where $x$ is the distance between atoms, $D$ is the well depth related
to the molecule dissociation energy, and $\alpha$ is an inverse length
parameter related to the curvature of the potential at the origin. As
such, this potential has been frequently used as an empirical model
for anharmonic interactions in the study of a large variety of
physical systems and conditions, including the rotating vibrational
states of diatomic molecules \cite{Dunham1932}, the adsorption of
atoms and molecules by solid surfaces \cite{LJ1935}, and the
deformation of cubic metals \cite{Girifalco1959}. Figure~1 shows that,
according to this potential, the energy difference between levels
gradually decreases as the level number $n$ increases, till it reaches
zero, where no more bounded states are allowed. 

Variants of the Morse potential have also been utilized to investigate
the physical behavior of more complex materials. As a typical example,
the state-of-the-art for atomistic modeling of semiconductor surfaces
and interfaces is the Abel-Tersoff potential
\cite{Abel1985,Tersoff1986} that has the form of a Morse pair
potential, but with the bond-strength parameter depending on its local
environment. Also in the investigation of thermal denaturation of
double-stranded DNA chains, the Morse potential has been successfully
applied to model hydrogen bonds connecting two bases in a pair
\cite{Gao1984a,Gao1984b,Peyrard1989,Theodorakopoulos2000}.

It is the purpose of this Letter to show that the Morse potential
emerges naturally as an effective interaction when a particle is
subjected to a harmonic potential in a contracted space. This
physical situation is substantiated here in terms of the following
quantum operator for non-additive translation
\cite{Costa2011,Mazharimousavi2012}:
\begin{equation}
\label{eq_def}
\langle x|U_{\gamma}(\varepsilon)|\psi \rangle=\psi(x+\varepsilon(1+g(\gamma x))),
\end{equation}
where there is no restriction on $g(\gamma x)$.  The action of
$U_{\gamma}(\varepsilon)$ on the bra vector $\langle x|$ can therefore
be expressed as $\langle x|U_{\gamma}(\varepsilon) =
(U_{\gamma}(\varepsilon)^{\dagger} |x \rangle)^{\dagger} = \langle
x+\varepsilon(1+g(\gamma x))| $. For infinitesimal transformations we
obtain that,
\begin{eqnarray}
\label{rep}
U_{\gamma}(\delta x) = {\mathbf 1}+ \frac{i}{\hbar}p_{\gamma}\delta x,
\end{eqnarray}
and
\begin{eqnarray}
\label{prep}
\frac{i}{\hbar}\langle x|p_{\gamma}|\psi \rangle = (1+g(\gamma x))\frac{d}{dx}\langle x|\psi \rangle,
\end{eqnarray}
where $p_{\gamma}$ is the momentum operator. Considering the
particular case where the function $g(\gamma x)=\gamma x$ and a finite
displacement $a$, we can rewrite Eq.~(\ref{eq_def}) as,
\begin{eqnarray}
\label{eq_finite}
\langle x |U_{\gamma}(a)|\psi \rangle = \psi\left( xe^{\gamma a}+ 
\frac{e^{\gamma a}-1}{\gamma}\right),
\end{eqnarray}
from which one can immediately recognize the action of the
dilation/contraction operator $xd/dx$. Moreover, as defined in
Eq.~(\ref{eq_def}), the non-additive operator $U_{\gamma}(a)$
corresponds to the infinitesimal generator of the $q$-exponential
function \cite{Tsallis2004}, 
\begin{equation}
 \label{q-exp}
 \exp_{q}(a) \equiv (1+(1-q)a)^{1/(1-q)}
\end{equation}
with $q=1-\gamma$. Equation~(\ref{q-exp}) represents a fundamental
mathematical definition for the generalized thermostatistics of
Tsallis and its applications
\cite{Tsallis2009,Adib2003,Andrade2002,Hasegawa2009,Andrade2010,Hanel2011,Nobre2011,Nobre2012}.

At this point, it is important to state that the momentum operator
$p_{\gamma}$ is Hermitian with regard to the following scalar product:
\begin{eqnarray}
\label{eq_sp}
(\psi , \phi) = \int \frac{dx}{1+ \gamma x} \psi ^*(x) \phi (x),
\end{eqnarray}
where the range of integration shall depend on the specific boundary
conditions of the system under investigation. Equation~(\ref{eq_def})
implies that the action of $U_{\gamma}(a)$ is unitary. Indeed, the
measure of integration $dx/(1+\gamma x)$ is invariant under the action
of the transformation, $x \rightarrow y= xe^{\gamma a}+ (e^{\gamma
  a}-1)/\gamma$, and the decomposition of the unit operator takes the
form,
\begin{equation}
 \label{unit}
 {\mathbf 1}=\int \frac{dx}{1+ \gamma x} |x\rangle \langle x|.
\end{equation}
The equation of motion for a particle in the $x$-representation of
this dilated/contracted space corresponds to a time-dependent
Schr\"odinger-like equation in the form,
\begin{equation}
\label{Schrodinger1}
i\hbar \frac{\partial}{\partial t}\psi(x,t)=H \psi(x,t),
\end{equation}
where the Hamiltonian operator is $H=\hat{p}_\gamma^2/2m+V(x)$, and
the modified momentum operator can be written for short as
$\hat{p}_{\gamma}=-i\hbar D_{\gamma}$, with $D_{\gamma}\equiv(1+\gamma
x) d/dx$ being a deformed derivative in space.
Equation~(\ref{Schrodinger1}) can then be rewritten as,
\begin{align}
\label{Shrodinger2}
i\hbar \frac{\partial}{\partial t}\psi(x,t) &
=-\left(\frac{\hbar^2}{2m}\right)D_\gamma^2 \psi(x,t)+V(x)\psi(x,t),
\end{align}
or, more explicitly, as
\begin{align}
\label{Shrodinger3}
i\hbar\frac{\partial}{\partial t}\psi(x,t)&=
-(1+\gamma x)^2\frac{\hbar^2}{2m}\frac{\partial^2}{\partial x^2}\psi(x,t)-
\nonumber\\&\gamma(1+\gamma x)\frac{\hbar^2}{2m}\frac{\partial}{\partial x}\psi(x,t)
+V(x)\psi(x,t).
\end{align}
However, it is more convenient to express Eq.~(\ref{Shrodinger3}) in
terms of a simple change of variables, as suggested by the
decomposition of the unit operator Eq.~(\ref{unit}). More precisely,
we can define a variable $\eta$ through the differential equation
$d\eta/dx=1/(1+\gamma x)$ with boundary condition $\eta(0)=0$, whose
solution is
\begin{equation}
\label{transf}
\eta=\frac{\ln(1+\gamma x)}{\gamma}. 
\end{equation} 
From this transformation, it is important to notice that the
``canonical coordinate'' can be written as
$x=(\exp(\gamma\eta)-1)/\gamma$.  In this way, the finite interval
$[0,L]$ for $x$ corresponds to $[0,{\bar L}]$, with ${\bar
  L}=\ln(1+\gamma L)/\gamma$, for the variable $\eta$.
Equation~(\ref{Shrodinger3}) rewritten in terms of the new variable
$\eta$ becomes,
\begin{equation}
i\hbar\frac{\partial}{\partial t}{\phi}(\eta,t)=
-\frac{\hbar^{2}}{2m}\frac{\partial^{2}}{\partial\eta^{2}}{\phi}(\eta,t)+
{V}_{\rm eff}(\eta){\phi}(\eta,t),
\end{equation}
where ${\phi}(\eta,t)=\psi(x(\eta),t)$ and $V_{\rm eff}(\eta)=V(x(\eta))$.
Assuming that,
\begin{equation}
\phi(\eta,t)=\Phi(\eta)\exp{(-iEt/\hbar)},
\end{equation}
the transformation (\ref{transf}) leads us back to a more familiar
version of the time-independent Schr\"odinger equation, 
\begin{equation}
E{\Phi}(\eta)=-\frac{\hbar^{2}}{2m}\frac{d^{2}}{d\eta^{2}}{\Phi}(\eta)+
V_{\rm eff}(\eta){\Phi}(\eta),
\end{equation}

If we now consider the problem of a standing wave in a null potential,
$V_{\rm eff}(\eta)=0$, it follows that the standard form of the plane wave
solution is recovered in terms of the transformed variable $\eta$,
namely $\Phi(\eta)=e^{\pm ik\eta}$. In the archetypal case of a
harmonic oscillator, $V(x)=\frac{1}{2}m\omega^2x^2$, the transformed
effective potential becomes,
\begin{equation}
\label{Morse2}
{V}_{\rm eff}(\eta)=\frac{m\omega^{2}}{2\gamma^{2}}(e^{\gamma\eta}-1)^{2}, 
\end{equation}
where $\omega$ is the frequency of the oscillator. Strikingly, by
identifying $D\equiv m\omega^{2}/2\gamma^{2}$ and $\alpha\equiv
-\gamma$, we conclude that Eq.~(\ref{Morse2}) corresponds exactly to
the expression (\ref{Morse}) for the Morse potential. To the best of
our knowledge, this is the first time that a connection based on
fundamental quantum principles is provided between this potential,
which has been widely utilized as a consistent description for the
vibrational structure of diatomic molecules, and the harmonic
potential. A physical interpretation for this correspondence can be
made in terms of a position-dependent (effective) mass induced due to
the presence of a material body in the system \cite{Costa2011} as, for
example, the case of electrons propagating through abrupt interfaces
in semiconductor heterostructures
\cite{BenDaniel1966,Roos1983,Bastard1992,Serra1997,Cavalcante1997}.

The wave function solution for the quantum Morse oscillator (QMO) has 
been previously determined as \cite{Morse1929},
\begin{equation}
\Phi_n(z)=A_nz^{s}e^{-\frac{1}{2}z}L_n^{2s}(z),
\end{equation} 
where $z=\nu e^{\gamma \eta}$,
$\nu\equiv{2m\omega}/{\gamma^{2}\hbar}$, $s=\nu/2-n-1/2$,
$L_n^{2s}(z)=\left(z^{-2s}e^{z}/n!\right)d^n\left(e^{-z}z^{n+2s}\right)/dz^n$
is the generalized Laguerre polynomial \cite{Arfken2005}, and the
normalization constant is given by,
\begin{equation}
A_n=\sqrt{\frac{n!}{\Gamma(\nu-n)}}.
\end{equation}
\begin{figure} 
  \epsfig{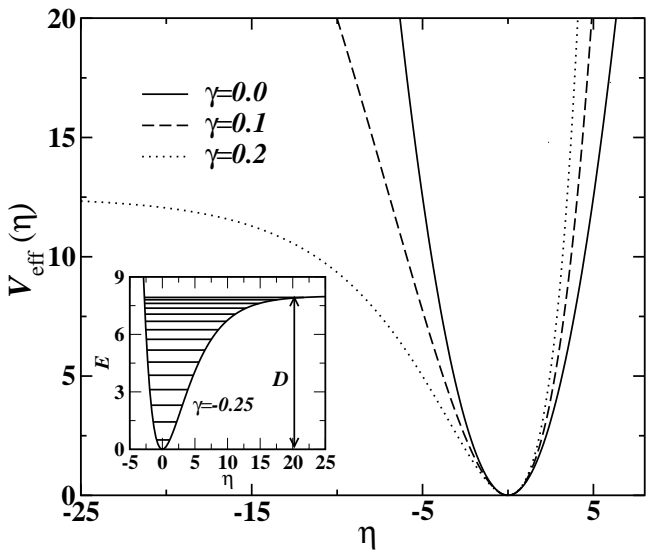}
  \caption{The effective potential given by $V_{\rm eff}(\eta) =
    D(1-e^{\gamma\eta})^{2}$ for $\gamma=0$ (solid line), $\gamma=0.1$
    (dashed line), and $\gamma=0.2$ (dotted line). Here we use
    $\omega=1$ and adopt atomic units, namely $\hbar=m=1$. As
    shown, the anharmonicity of the potential depends on $\gamma$. The
    energy levels for $\gamma=-0.25$ are shown in the inset. The
    parameter $D$ corresponds to the dissociation energy of the
    diatomic molecule.}
  \label{levels}
\end{figure}
The energies for the QMO can then be calculated as,
\begin{equation}
\label{energy_Morse}
E_n=\hbar\omega\left(n+\frac{1}{2}\right)-\frac{\hbar^{2}\gamma^{2}}{2m}
\left(n+\frac{1}{2}\right)^{2}.
\end{equation}
This expression clearly indicates that, in the limit of small values
of $\gamma$, the QHO spectrum is recovered. One should also note that
the energy levels,
\begin{equation}
\Delta E=E_{n+1}-E_n=\hbar\omega-\frac{\hbar^{2}\gamma^{2}}{m}(n+1),
\end{equation}
valid for $0\leq n \leq \lfloor m\omega/\gamma^{2}-1 \rfloor$, are no
longer equally spaced, and approach zero depending on $\gamma$ and the
quantum number $n$. In the main plot of Fig.~1 we show the form of the
potential (\ref{Morse2}) for different values of $\gamma$, while in
the inset the energy levels are depicted for $\gamma=-0.25$. The
potential is symmetric in $\gamma$, i.e., $V(\gamma)=V(-\gamma)$. The
energy difference between levels decreases as the quantum number
increases.
 
Finally, we show that an uncertainty relation for the QMO can be
disclosed in the framework of our theoretical approach. In order to do
this, expected values for $x$ and $p$ must be calculated. Considering
that the following correspondence holds between expected values of a
given operator $\hat{O}$ in the $x-$space and $\eta-$space:
\begin{equation}
\left\langle \hat{O}_{x}\right\rangle _{x}=
\left\langle \hat{O}_{\eta}\right\rangle _{\eta}=
\int d\eta\psi_n(\eta)^* O_{\eta}\psi_n(\eta),  
\end{equation}
and using a procedure similar to the one adopted in
Ref.~\cite{Zuniga1998}, we obtain,
\begin{align}
  \left\langle x\right\rangle _{n,x}&=-\frac{\gamma\hbar}{m\omega}\left(n+\frac{1}{2}\right),\\
  \left\langle p\right\rangle _{n,x}&=0,\\
  \left\langle x^{2}\right\rangle _{n,x}&=\frac{\hbar}{m\omega}\left(n+\frac{1}{2}\right),\\
  \left\langle p^{2}\right\rangle
  _{n,x}&=m\hbar\omega\left(n+\frac{1}{2}\right)\left[1-\frac{\hbar\gamma^2}{m\omega}\left(n+\frac{1}{2}\right)\right],
\end{align}
so that the uncertainty relation for the QMO can be written as,
\begin{equation}
\label{uncertainty_Morse}
\Delta x\Delta p=\hbar\left(n+\frac{1}{2}\right)\left[1-\frac{\gamma^2\hbar}{m\omega}
\left(n+\frac{1}{2}\right)\right].
\end{equation}
For the ground state, this expression is identical to the one
previously obtained in Ref.~\cite{Costa2011}, namely $\Delta x\Delta
p\ge (\hbar/2)(1+\gamma\langle x\rangle)$. The uncertainty in the QMO
is therefore lower than in the QHO due to the anharmonicity of the
Morse potential. Note that in both expressions for energy
(\ref{energy_Morse}) and uncertainty (\ref{uncertainty_Morse}) the
parameter $\gamma$, which defines whether the space is contracting
($\gamma<0$) or dilating ($\gamma>0$), appears squared. Consequently,
for the harmonic potential in a deformed space, the energy is
symmetric under contraction or dilation.
 
It is interesting to note that the Morse potential can be associated
to a two-dimensional harmonic oscillator in polar coordinates for
$\omega=1$, and considering atomic units, $\hbar=m=1$. In fact, it is
this relationship that allows one to solve the QMO using ladder
operators \cite{Montemayor1983}, and study this problem through
supersymmetry \cite{Plastino1999}. As a consequence, our approach also
provides a conceptual basis for the map between a 1D-QHO and a 2D-QHO
in polar coordinates. This mapping is the result of the new
commutation relation for $\hat{x}$ and $\hat{p}$ generated through the
translation operator introduced in \cite{Costa2011} and used here.

Previous studies have focused on plausible modifications on the
position momentum \cite{Kempf1995,Hinrichsen1996,Kempf1997}, so that a
minimum length and momentum could be defined for quantum theory. In
particular, Quesne {\it et al.} \cite{Quesne2007} have shown that, if
some special generalized deformed commutation relations are employed
(e.g., $[-e^{-x},p]=i[e^{-x}+\beta p^2]$), the Morse potential can be
obtained as an effective potential of the theory. In all these
studies, however, modified commutation relations are introduced in an
{\it ad hoc.} manner, i.e., they are not obtained from a first
principle mechanism, like the non-additive translation operator
employed in this work. Another important point here is that the
parameter $\gamma$, responsible for the dilation/contraction in the
translation, corresponds exactly to the minus value of the $\alpha$
parameter for the Morse potential. In the particular case of the
Hydrogen molecule, for example, the numerical value of this parameter
is $\gamma=-\alpha=-1.4$ $a.u$.

In summary, we have shown that a one-dimensional harmonic potential in
a space deformed by the action of the operator defined as in
Eq.~(\ref{eq_def}) can be equivalent to the Morse potential in a
regular space. As the particle travels in a different way in the
deformed space, it feels the harmonic potential as a Morse potential
in regular space. This equivalence is achieved when $g(\gamma
x)=\gamma x$, namely for the case in which the translation occurs as
a contraction in a deformed space, $\gamma<0$. Such a physical
framework has perfect analogy with the behavior of a
position-dependent (effective) mass particle on a non-homogeneous
substrate, typified by electrons moving through abrupt interfaces in
semiconductor heterostructures. In this particular situation, the
anharmonic feature of the Morse potential emerges naturally from the
standard quantum harmonic oscillator. We thus conclude that our study,
based on a non-additive translation operator, provides a first
principle explanation for anharmonic properties.
 
\acknowledgments{We thank the Brazilian Agencies CNPq, CAPES, FUNCAP
  and FINEP, the FUNCAP/CNPq Pronex grant, the National Institute of
  Science and Technology for Complex Systems in Brazil, the Centre for
  Advanced Study (CAS) in Norway, and the Norwegian Research Council
  for financial support.}

\end{document}